\documentclass[prb,aps,10pt,twocolumn]{revtex4}
\begin{document}
\title{Effects of the electron-phonon coupling near and within the insulating Mott phase}
\author{C.A. Perroni, V. Cataudella, G. De Filippis, and V.
Marigliano Ramaglia}
\affiliation{Coherentia-INFM  and Dipartimento di Scienze Fisiche, \\
Universit\`{a} degli Studi di Napoli ``Federico II'',\\
Complesso Universitario Monte Sant'Angelo,\\
Via Cintia, I-80126 Napoli, Italy}
\date{\today}

\begin{abstract}

The role of the electron-phonon interaction in the
Holstein-Hubbard model is investigated in the metallic phase close
to the Mott transition and in the insulating Mott phase. The model
is studied by means of a variational slave boson technique. At
half-filling, mean-field static quantities are in good agreement
with the results obtained by numerical techniques. By taking into
account gaussian fluctuations, an analytic expression of the
spectral density is derived in the Mott insulating phase showing
that an increase of the electron-phonon coupling leads to a
sensitive reduction of the Mott gap through a reduced effective
repulsion. The relation of the results with recent experimental
observations in strongly correlated systems is discussed.

\end{abstract}

\maketitle

\newpage
\section{Introduction}
The understanding of the properties of Mott insulators such as
$CaTiO_3$ and $V_2O_3$ represents a long-standing problem. As
suggested by Mott, \cite{mott} the strong Coulomb repulsion among
the electrons can cause a metal-insulator transition opening a gap
in the density of states that is usually known as Mott gap. The
correlation-driven metal-insulator transition is often
investigated within the framework of the Hubbard lattice
Hamiltonian which takes into account the electron interaction
through the on-site repulsion term $U$. As developed by Hubbard,
\cite{hubb} two subbands generically called lower and upper
Hubbard bands separated by the Mott gap of the order of the energy
$U$ arise in the excitation spectrum for large enough repulsion.

The correlation effects are important not only on the insulating
side but also on the metallic phase where a great deal of insight
has been obtained by using the Gutzwiller wave function.
\cite{brink}  A progress toward the understanding of the
metal-insulator transition has been made by the reformulation of
the problem in terms of slave bosons, \cite{kotl} which reproduces
the Gutzwiller approximation at the saddle-point level and allows
to study the effects of gaussian fluctuations in the metallic
\cite{rasul} and in the insulating phase. \cite{castel} In this
state the resulting single-particle spectral function consists of
two broad incoherent contributions reminiscent of the lower and
upper Hubbard bands. An overall description of the Mott transition
in the Hubbard model can be obtained by means of the dynamical
mean field theory ($DMFT$) that is exact in the limit of infinite
dimension. \cite{georges} The study of the infinite dimensional
model has essentially confirmed the validity of the description
obtained when slave boson fluctuations about the mean-field ($MF$)
are considered.

Although the Hubbard Hamiltonian captures the fundamental
properties of systems near the Mott transition, it does not take
into account the role of the lattice degrees of freedom. Actually,
the presence of strong electron-phonon ($el-ph$) interactions has
been pointed out in several systems, such as cuprates,
\cite{cupr1,cupr} manganites, \cite{mang} and $V_2O_3$.
\cite{vanad}. In addition to the on-site repulsion, the most
natural model incorporates a short-range interaction of the
electrons with local phonon modes of constant frequency
$\omega_0$. \cite{holstein} This model is described by the
Holstein-Hubbard Hamiltonian $H$

\begin{eqnarray}
H&=&-t \sum_{<i,j>, \sigma} c^{\dagger}_{i \sigma} c_{j \sigma}+ U
\sum_{i} n_{i \uparrow} n_{i \downarrow}+ \nonumber \\
&& \omega_0 \sum_{i} a^{\dagger}_{i}a_{i}+ g \omega_0 \sum_{i}
n_{i} \left( a_{i}+a^{\dagger}_{i} \right). \label{1ra}
\end{eqnarray}
Here  $t$ is the electron transfer integral between nearest
neighbor sites $<i,j>$ of a $d$-dimensional simple cubic lattice,
$c^{\dagger}_{i \sigma} \left( c_{i \sigma} \right)$ creates
(destroys) an electron with spin $\sigma$ at the site $i$  and
$n_{i}=\sum_{\sigma} n_{i \sigma}= \sum_{\sigma} c^{\dagger}_{i
\sigma} c_{i\sigma} $ is the local density operator. In
Eq.(\ref{1ra}) $ a^{\dagger}_{i} \left( a_{i} \right)$ is the
creation (annihilation) phonon operator at the site i, and the
parameter $g$ represents the coupling constant between electrons
and local displacements. The dimensionless parameter $\lambda= g^2
\omega_0/W$, with $W=2 d t$ bare electron half bandwidth, is
typically used to measure the strength of the $el-ph$ interaction
in the adiabatic regime characterized by small values of the ratio
$\gamma=\omega_0/t$. Within this regime the Coulomb repulsion is
found to dominate the properties of the metallic phase also with a
sizable $el-ph$ coupling. \cite{hew} Furthermore, there is a very
little softening of the phonon frequency on the approach to the
Mott transition since the Hubbard term $U$ suppresses charge
fluctuations. \cite{hew} Actually within the Mott phase the
spectral density shows phonon side bands whose peaks are separated
by multiples of the bare frequency $\omega_0$. \cite{fes} Finally,
close to the Mott transition at finite density, an intermediate
$el-ph$ coupling leads to the phase separation between a metal and
an insulator. \cite{mil}

In this paper the study of the Holstein-Hubbard model focuses on
the role of the $el-ph$ interaction in modifying the physical
properties of the electrons close to the metal-insulator
transition and in the insulating Mott phase. The starting point of
the approach is the generalized Lang-Firsov transformation
\cite{lang} $U=e^V$, with
\begin{equation}
V=g \sum_{i} [<n_i>+f_i (n_i-<n_i>)](a_{i}- a^{\dagger}_{i}),
\label{3r}
\end{equation}
where the parameters $f_i$ take into account the polaronic local
density fluctuations which couple to the lattice distortions. The
next step is the functional-integral representation in terms of
the four-slave bosons $e_i$, $p_{i \sigma}$, and $d_i$ obtained by
imposing the equivalence with the original model through the
Lagrange multipliers $\lambda_i^{(1)}$ and $\lambda_{i
\sigma}^{(2)}$. \cite{kotl} First we consider the $MF$ solution at
half-filling, then slave boson gaussian fluctuations on the top of
the $MF$ finding that the resulting Mott gap is sensitively
reduced with increasing $\lambda$ since it is determined by the
$el-el$ repulsion renormalized by the effects of the $el-ph$
coupling. Finally the relation of the results with recent
experimental observations in strongly correlated systems is
discussed.

The $MF$ solution in the paramagnetic homogeneous phase is
obtained by replacing the Bose fields with their mean values
($\langle e_i \rangle=e_0$, $\langle d_i \rangle=d_0 $, $\langle
p_{i \sigma} \rangle=p_0$) and by assuming $f_i=f$,
$\lambda_i^{(1)}=\lambda_0^{(1)}$, and $\lambda_{i
\sigma}^{(2)}=\lambda_{0}^{(2)}$. \cite{cataud} The resulting
mean-field Hamiltonian is characteristic of free fermion
quasi-particles and phonons and the mean double occupation is
controlled by $U_{eff}=U+2 g^2 \omega_0 (f^{2}-2f)$.

At half-filling the $MF$ energy per site $u$ is given by a simple
functional depending on $f$ and $d_0$
\begin{equation}
u=q e^{- f^2 g^2} {\bar \varepsilon}+U_{eff} d_{0}^2- g^2
\omega_0, \label{8re}
\end{equation}
with $q=8 d_{0}^2-16 d_{0}^4$ and ${\bar \varepsilon}$ mean bare
kinetic energy. The Mott phase is the insulating state for large
$U_{eff}>0$ characterized by $q=0$, $d_0=0$, and the energy per
site equal to $- g^2 \omega_0$ (characteristic energy of the limit
$U/t=\infty$). When $U_{eff}$ becomes negative, a on-site
bipolaron solution sets in corresponding to $d_{0}^2=0.5$, $f=1$,
and energy $u=U/2-2 g^2 \omega_0$.

In Fig. 1 the phase diagram at half-filling is reported for
$\gamma=0.2$ in the three-dimensional case. By analyzing the
behavior of $d_0$, it is found that the transition to $MI$ is
always found to be second order, that to the bipolaronic insulator
($BI$) is first order, finally that between $MI$ and $BI$ is again
discontinuous in excellent agreement with the results derived by
the $DMFT$ solution. \cite{hew} In particular we notice that, with
increasing $\lambda$, the line separating the $M$ and $MI$ phases
shifts to values of $U$ larger than $U_c$, the critical value in
absence of the $el-ph$ coupling. Actually the Mott transition is
influenced by the $el-ph$ interaction since it is $U_{eff}$ that
governs the transition and it becomes smaller with increasing
$\lambda$. Therefore, the condition $U_{eff} \simeq U_c$,
characteristic of a transition driven by the $el-el$ interaction,
implies that the transition occurs for larger values of the bare
$U$. Within the $MF$ approach the interplay between $el-el$ and
$el-ph$ interactions in affecting the Mott phase is essentially
linked to the value of the parameter $f$ that, at fixed
adiabaticity ratio and $\lambda$, is weakly decreasing with
increasing $U$ and it is of the order of $\gamma/4 d$ in the
adiabatic regime near the Mott transition. The transition line
between $M$ and $MI$ phase is given by $\lambda \simeq (U-U_c)/(2
f W) \simeq (U-U_c)/\gamma t$. Therefore, as shown in the inset of
Fig. 1, the dependence of the Mott transition line on $\lambda$
becomes also more pronounced with increasing the adiabaticity
ratio $\gamma$. In the atomic limit ($\gamma=\infty$) we recovers
the exact solution without metallic phase with $f=1$ and $U_c=0$
(dotted line in Fig. 1 corresponding to $\lambda=U/2W$).

While the Mott transition is driven by the growth of the spin
susceptibility, the transition from $M$ to $BI$ is characterized
by the enhancement of the charge fluctuations inducing a decrease
of the effective repulsion. In Fig. 2 we report the $MF$ results
for the spectral weight $Z$ at the Fermi energy (equal to
$m/m^{*}$) and the local magnetic moment $M=<(M_{i}^{z})^2>$
stressing the effects of the $el-ph$ interaction. Within the $MF$
approach we simply have $Z=q e^{- f^2 g^2}$ and $M=1-2d_0$. Far
from the Mott transition ($U/U_c$ small) $Z$ decreases with
increasing the $el-ph$ coupling as expected for any localizing
interaction. However, near the Mott phase, the effects due to the
reduction of $U_{eff}$ become more marked and are able to induce
the enhancement of $Z$ with increasing $\lambda$ that, again, is
in good agreement with $DMFT$ results. \cite{hew} In Fig. 2 we
also show that the $M$ phase is reduced in comparison with its
value at $\lambda=0$ for any ratio $U/U_c$ implying that the
double occupation $d_0$ increases by approaching the $BI$ phase.
Therefore, while $M$ increases as function of $U$, it decreases as
function of $\lambda$.

The $MF$ solution can be readily generalized at densities
different from half-filling. Within the $MF$ approach some of us
have shown that the interplay of strong local $el-el$ and $el-ph$
interactions can push the system toward a phase separation between
states characterized by different lattice distortions.
\cite{cataud} The phase coexistence occurs for intermediate values
of the $el-ph$ coupling and its relevance within the
Hubbard-Holstein model has been confirmed also by $DMFT$ works.
\cite{mil}

The task of including charge fluctuations described by the $e$ and
$d$ fields is simplified in the Mott phase by the fact that at
$MF$ level the Bose fields $e$ and $d$ are not condensed
($e_0=d_0=0$), while $p_0=1/\sqrt 2$. \cite{castel} At gaussian
level the fluctuations of the $e$ and $d$ fields are actually
decoupled not only from those of the $p$ fields but also from the
phononic $a$ fields. Clearly the matrix elements of the
fluctuating fields $e$ and $d$ are renormalized by $el-ph$
contributions because of $U_{eff}$. The procedure of calculation
is the following: we start at densities different from
half-filling ($n<1$), then we calculate the quantities in the
limit $n\rightarrow 1^-$ for values of the parameters where the
system is in the Mott phase at half-filling. \cite{castel}

The inclusion of gaussian fluctuations allows to calculate the
one-particle spectral function $A(\omega)$ of the insulating Mott
phase in the limit $n\rightarrow 1^-$ (at $MF$ it is identically
zero). The incoherent contribution arises from the complicated
motion of a quasi-particle surrounded by the cloud of charge and
lattice excitations that it leaves behind. In the limit
$n\rightarrow 1^-$ it is possible to derive an analytic expression
of the spectral function $A(\omega)$ yielding at zero temperature
\begin{eqnarray}
A(\omega) &=& e^{-g^2 f^2} {\tilde A}(\omega)+
\\ \nonumber
&& e^{-g^2 f^2} \sum_{n=1}^{\infty} \frac{(g^2 f^2)^n}{n!}
\theta(-\omega-n\omega_0) {\tilde A}(\omega+n\omega_0)+
\nonumber \\
&& e^{-g^2 f^2} \sum_{n=1}^{\infty} \frac{(g^2 f^2)^n}{n!}
\theta(\omega-n\omega_0) {\tilde A}(\omega-n\omega_0), \label{21r}
\end{eqnarray}
with $\theta(x)$ Heaviside function and the function $ {\tilde
A}(\omega)$ akin to that obtained for the Hubbard model with
$U=U_{eff}$. \cite{castel}

The first term of Eq. (\ref{21r}) is the product of two
quantities, with $e^{-f^2 g^2}$ renormalization factor due to
$el-ph$ coupling. Through ${\tilde A}(\omega)$ this term is able
to reconstruct lower and upper Hubbard bands. The second and the
third term in Eq. (\ref{21r}) represent the contribution due to
the phonon replicas of hole- and particle-type, respectively.
\cite{perroni} In Fig. 3(a) we report the spectral density
$N(\omega)=A(\omega)/2 \pi$ (solid line) together with the
resulting first term of Eq. (\ref{21r}) (dotted line) and the
contribution from the phonon replicas (dashed line). We notice
that this last term provides a non negligible spectral weight to
the total spectral density at energies out of the gap of ${\tilde
A}(\omega)$. Therefore, the Mott gap of the system is determined
by the gap of the function ${\tilde A}(\omega)$ and it is simply
given by $\Delta=U_{eff} \xi$. As seen in Fig. 3(b), the reduction
of the gap with increasing $\lambda$ can be also of the order of
bare half bandwidth $W$. \cite{phase} Finally we stress that the
calculated gap is traceable to the difference of the chemical
potentials at $n=1$ and $n=1^{-}$, in analogy with the results of
the Hubbard model. \cite{castel} For $n=1$ we have $\mu(1)=U/2 -
2g^2 \omega_0 $, while for $n=1^-$ the chemical potential is
$\mu(1^-)=\mu(1)-U_{eff} \xi /2$. Since the system has
particle-hole symmetry, the gap $\Delta$ can be obtained as
$\Delta=2[\mu(1)-\mu(1^-)]=U_{eff} \xi $.

In Fig. 4 we report the difference between the gap at finite
$\lambda$ and that at $\lambda=0$ as a function of the
adiabaticity ratio. Since the attraction between the electrons
gets larger with increasing $\gamma$, the resulting Mott gap is
more reduced. However in the strong $el-ph$ coupling regime
($\lambda$ larger than 1) there is no dependence on the
adiabaticity ratio since the particles localized by the strong
correlation are small polarons. In fact in this regime the $MF$
solution at finite density is minimized for $f=1$ yielding the
effective interaction $U_{eff}=U-2 g^2\omega_0$ for any finite
value of adiabatic ratio. In the inset of Fig. 4 we show the
variation of the effective interaction $U_{eff}$ as a function of
the adiabaticity ratio making the comparison with the behavior of
the Mott gap. In the regime where $\lambda<1$, the quantity $\xi$
affects the magnitude of the gap and its dependence on the
adiabaticity ratio, while in the strong coupling regime $\xi
\simeq 1$ implying that the gap in units of $W$ is just
$U_{eff}/W=U/W-2\lambda$.

In this work we have seen that in the metallic phase close to the
Mott transition the spectral weight $Z$ is enhanced and the gap in
the insulating Mott phase is reduced as $\lambda$ is raised. These
behaviors can be related to some recent experimental and
theoretical studies in $V_2O_3$ and $Cr$-doped $V_2O_3$
\cite{vanad,mo,keller} where the Mott gap is unexpectedly small
and in the metallic phase the quasi-particle peak in the
photoemission spectrum has a significantly large weight in
comparison with that theoretically predicted. We suggest that the
inclusion of the $el-ph$ interaction could be able to partially
fill the discrepancies between the experimental observations and
the theoretical studies. Clearly the single orbital model is not
sufficient to fully explain the electronic and magnetic structure
of such systems, \cite{refero} so a proper multi-orbital theory
has to be considered in order to obtain a better agreement of the
theory with experiments. \cite{keller} The results due to the
reduction of the effective repulsion caused by the $el-ph$
coupling are valid when the lattice distortions are coupled to
charge fluctuations like in the model of Eq.(\ref{1ra}). However
for general models the issue is delicate since there are
interactions such as the Jahn-Teller coupling for which the
phonon-mediated attraction could be even diminished by strong
correlations. \cite{han}

Finally we can argue the modifications of the spectral properties
at densities near the $MI$ phase and at half-filling just under
the edge of localization. Clearly there is a finite spectral
weight at the Fermi energy where quasi-particle states begin to
appear. Therefore, in addition to the incoherent contribution at
high energy, the spectral density presents also the coherent term.
Close to the metal-insulator transition the coherent term could
not be strongly affected by the $el-ph$ coupling. In fact for the
combined effect of the strong correlation and $el-ph$ coupling the
quasi-particle band can be narrower than the phonon energy
$\omega_0$ implying the impossibility of the single phonon
scattering between the quasi-particles. Therefore near the Mott
transition the $el-ph$ coupling affects mainly the incoherent term
of the spectral density. This result is in agreement with recent
experiments on $Bi_2Sr_2CaCu_2O_{8+\delta}$ made using
angle-resolved photoemission spectroscopy. \cite{cupr} In fact it
has been found that the oxygen isotope substitution mainly
influences the broad high energy humps.

In conclusion, we have discussed the role of the $el-ph$
interaction in modifying the physical properties of the electrons
in the metallic phase close to the Mott transition and in the $MI$
phase. The approach to study the Holstein-Hubbard model has been
based on a variational slave boson technique that provides results
in agreement with $DMFT$. An analytic expression of the spectral
density is derived in the Mott phase showing that due to the
reduced effective repulsion the Mott gap decreases as the $el-ph$
coupling constant increases. In this paper we have mainly
discussed the phases without long-range order. The study of broken
symmetry phases is possible within the slave-boson formalism
\cite{kotl} and it is left for future work.

\section*{Figure captions}
\begin{description}

\item  {F1}
The phase diagram $U/W$ versus $\lambda$ at half-filling for
$\gamma=0.2$ in the three-dimensional case. The transition lines
separate the metallic state $M$ from the Mott insulator $MI$ and
the bipolaronic insulator $BI$. The dotted line is the locus where
$U- 2 g^2 \omega_0 =0 $. In the inset, the transition line between
the metallic and Mott insulating phase is shown for different
adiabaticity ratios $\gamma$.

\item  {F2}
The difference between the spectral weight $Z$ at the Fermi energy
(a) and the local magnetic moment $M$ at finite $\lambda$ (b) and
their respective values at $\lambda=0$ for some values of the
ratio $U/U_c$ in the three-dimensional simple cubic lattice.

\item  {F3}
(a) The renormalized density of states $N$ (solid line) together
with the dominant contribution $e^{-g^2 f^2}$ ${\tilde A}(\omega)$
(dotted line) and the term due to the phonon replicas (dashed
line) as function of the frequency $\omega$.

(b) The renormalized density of states $N$ for different values of
the $el-ph$ coupling constant $\lambda$ as a function of the
frequency $\omega$.

\item  {F4}
The difference between the gap at $\lambda=0$ and that at finite
$\lambda$ in units of the bare half bandwidth $W$ as function of
the adiabaticity ratio for different values of the $el-ph$
coupling. In the inset the effective repulsion $U_{eff}$ and the
Mott gap $\Delta$ as a function of the adiabaticity ratio
$\gamma$.

\end{description}


\begin{references}
\bibitem{phase} The spectral density is not shown for values of $\lambda$
close to 1 since the system shows phase separation at densities
different from half-filling.
\end{references}
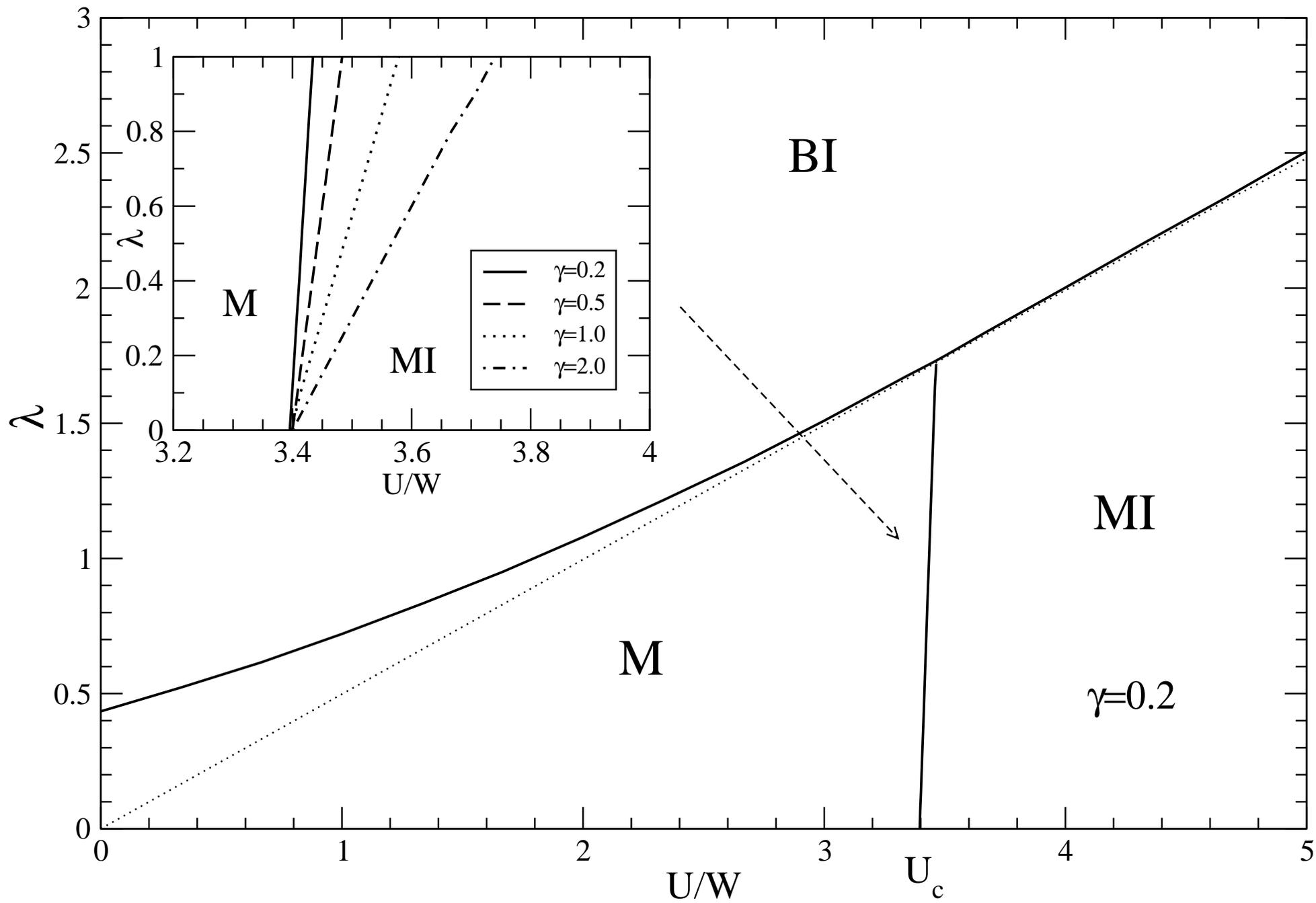
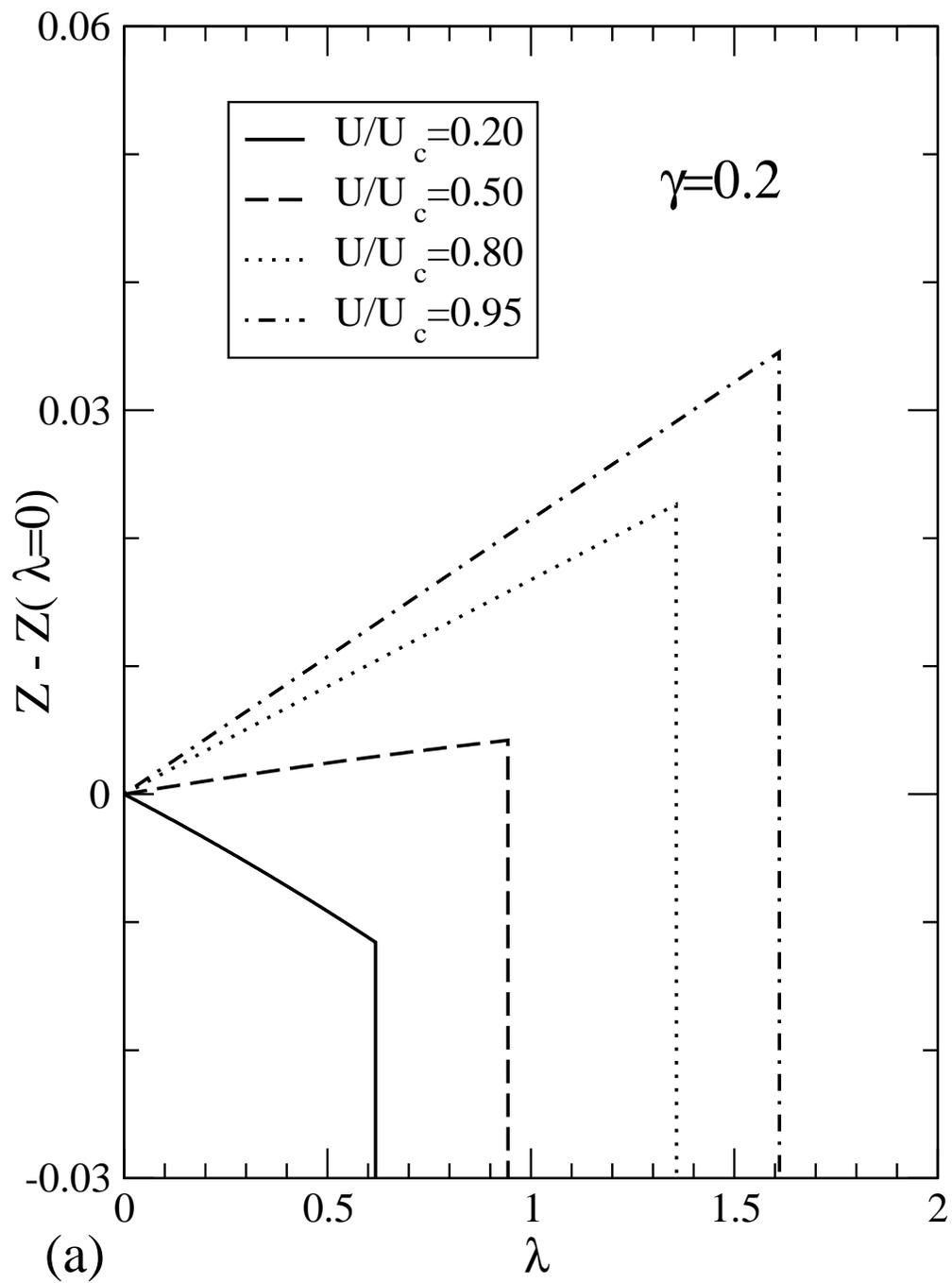
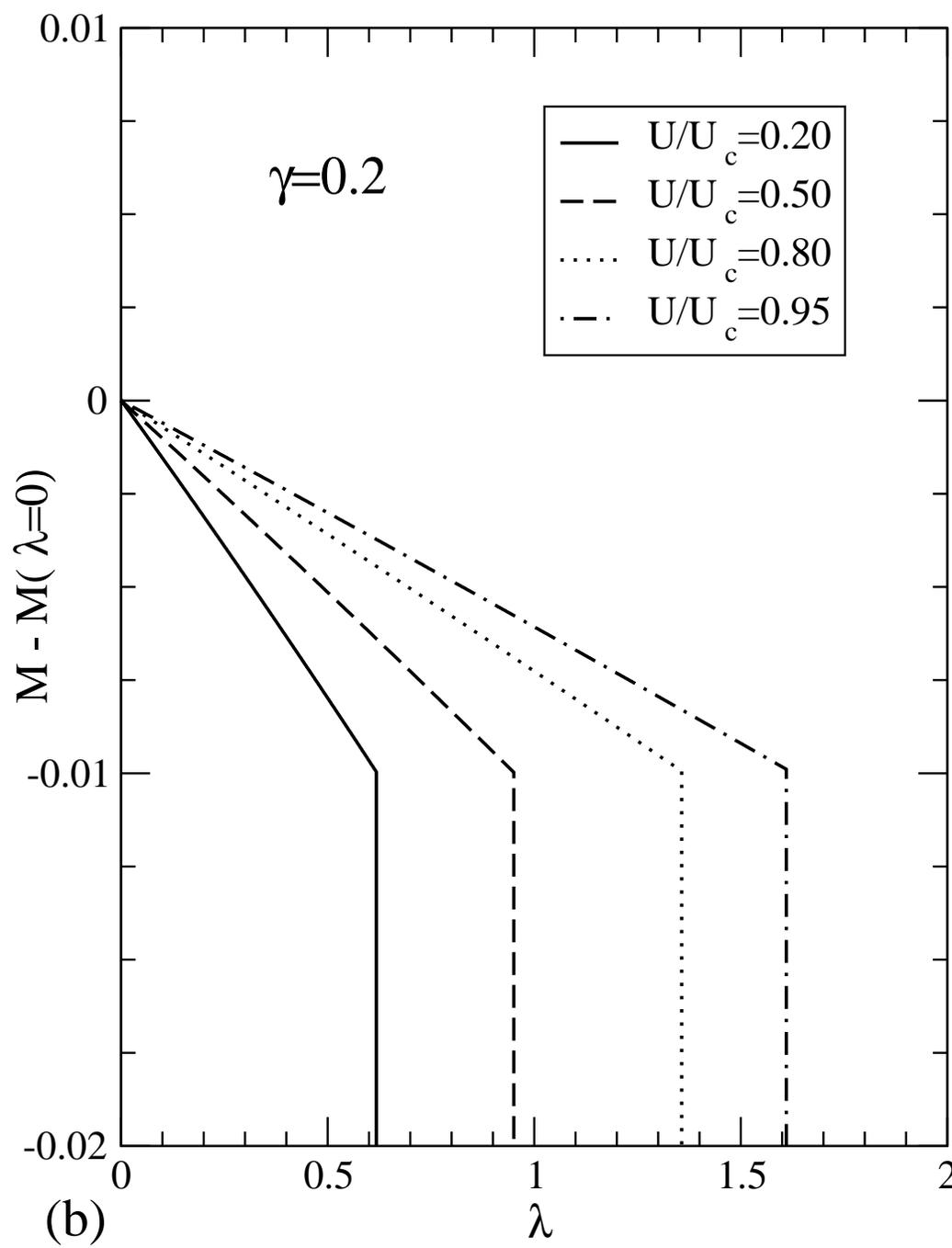
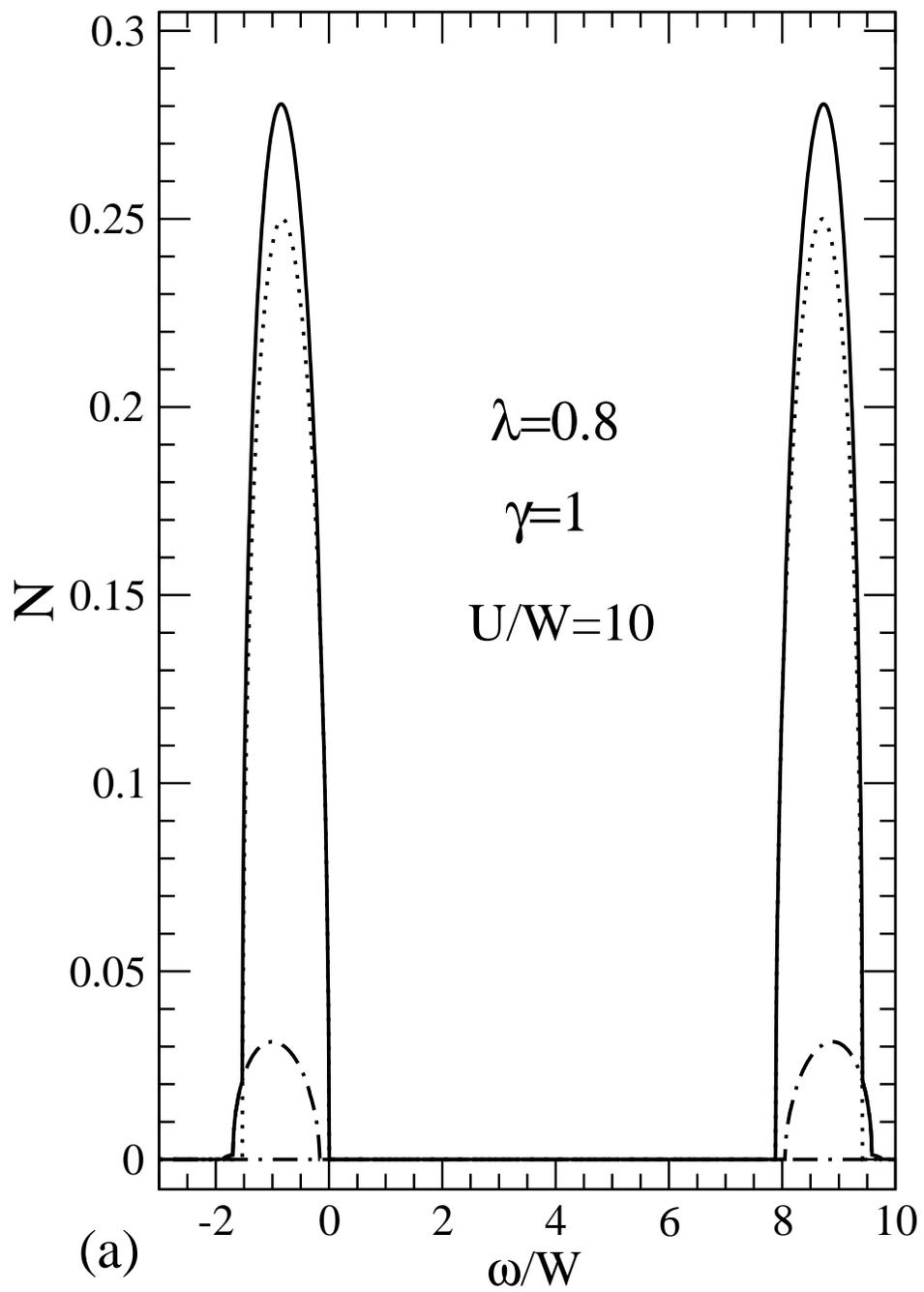
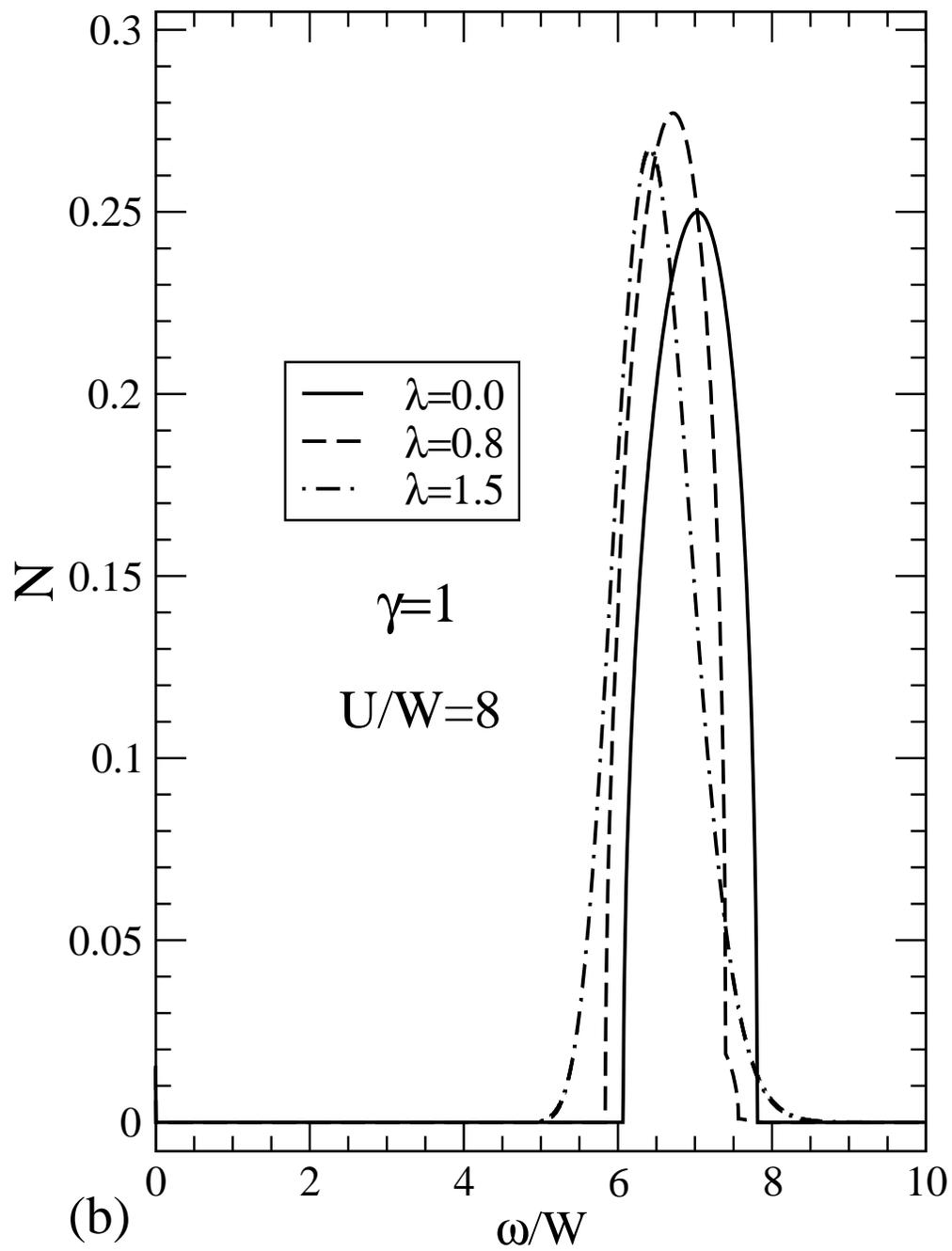
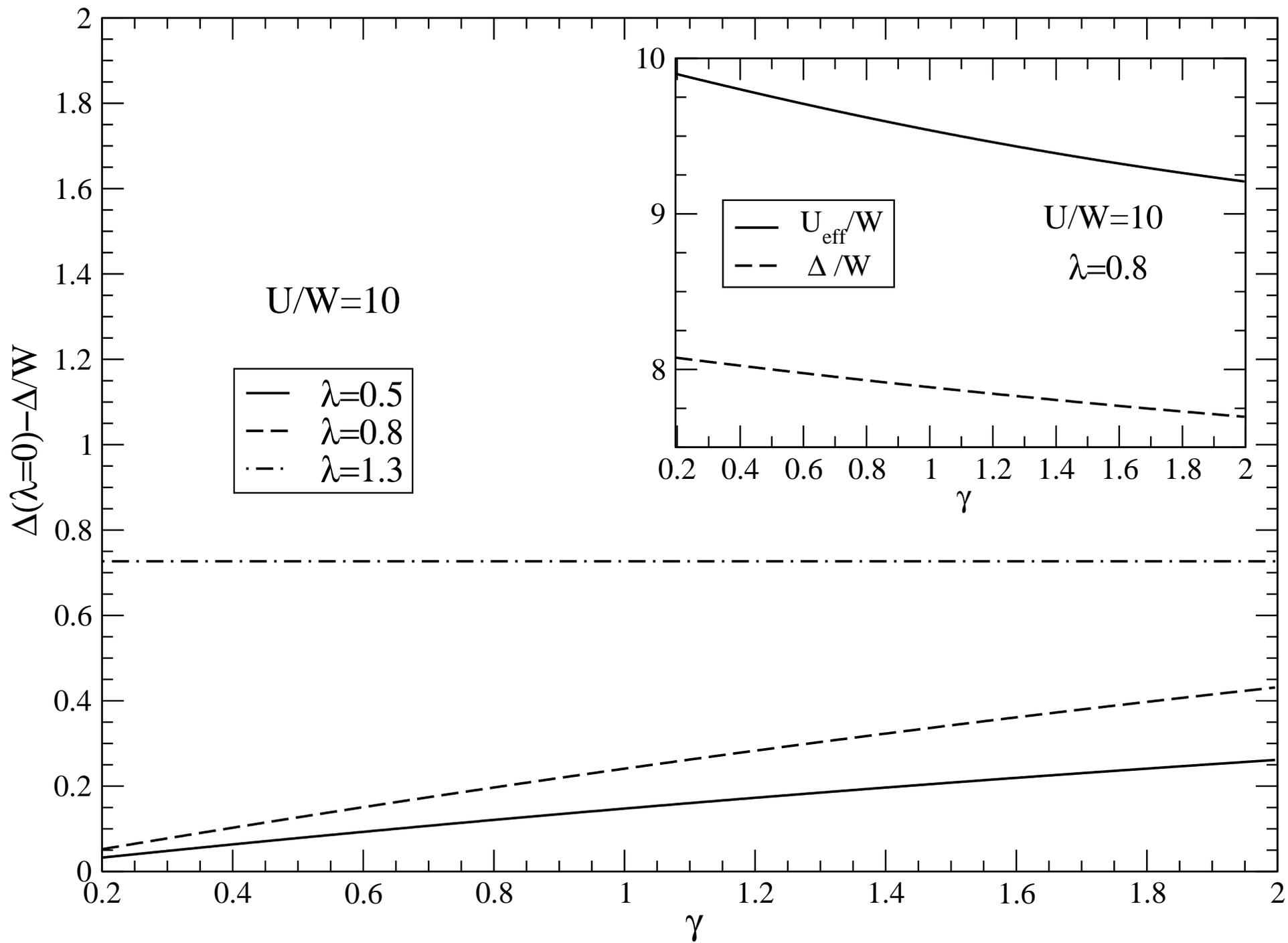
\end{document}